\documentclass{sig-alternate} 
\usepackage{mathptmx} % This is Times font

\newcommand{\ignore}[1]{}
\usepackage{fancyhdr}
\usepackage[normalem]{ulem}
\usepackage[hyphens]{url}
\usepackage{microtype}

\usepackage{amsmath,amssymb,amsfonts}
\usepackage{algorithmic}
\usepackage{graphicx}
\usepackage{textcomp}
\usepackage{xcolor}
\usepackage{courier}

\usepackage[bookmarks=true,breaklinks=true,letterpaper=true,colorlinks,linkcolor=black,citecolor=blue,urlcolor=black]{hyperref}

\fancypagestyle{firstpage}{
  \fancyhf{}

  \fancyhead[C]{\normalsize{ Accepted in \emph{The Third Data Prefetching Championship (DPC3)}, held in conjunction with \emph{ISCA 2019}}}
 \fancyfoot[C]{\thepage}
}  

\title{Pangloss: a novel Markov chain prefetcher}

\author{
Philippos Papaphilippou, Paul H. J. Kelly, Wayne Luk\\
\large
\textit{Department of Computing, Imperial College London, UK}\\
{
\{pp616, p.kelly, w.luk\}@imperial.ac.uk}
}

\usepackage[backend=bibtex,style=ieee,sorting=none]{biblatex}
\bibliography{Pangloss}

\begin{document}
\maketitle
\thispagestyle{firstpage}
\pagestyle{plain}

\begin{abstract} %300-word limit
We present Pangloss, an efficient high-performance data prefetcher that approximates a Markov chain on delta transitions. With a limited information scope and space/logic complexity, it is able to reconstruct a variety of both simple and complex access patterns. %, such as non-strided address progressions. 
This is achieved by a highly-efficient representation of the Markov chain to provide accurate values for transition probabilities. In addition, we have added a mechanism to reconstruct delta transitions originally obfuscated by the out-of-order execution or page transitions, such as when streaming data from multiple sources. Our single-level (L2) prefetcher achieves a geometric speedup of 1.7\% and 3.2\% over selected state-of-the-art baselines (KPCP and BOP). When combined with an equivalent for the L1 cache (L1 \& L2), the speedups rise to 6.8\% and 8.4\%, and 40.4\% over non-prefetch. In the multi-core evaluation, there seems to be a considerable performance improvement as well.

%mention IP, H/W friendliness

% In our milti-program evaluation the speeups are

%Our prefetcher is simplistic in regards to issuing prefetches, as it mainly follows the generated Markov model. 
% utilise the current knowledge on a wider scope, such as when experiencing the same offset transitions but in a different page or from different instruction.
\end{abstract}

%\begin{keywords}
%Markov chain, delta transitions, offset transitions, distance prefetching, data prefetcher
%\end{keywords}

\section{Introduction} %Motivation as a subsection?

Markov models have been used extensively in prior research for prefetching purposes, by estimating and utilising address transition probabilities for subsequent accesses. Distance prefetching is a generalisation of the common Markov model prefetchers \cite{ghb}, that uses deltas instead of addresses to build more general models (originally for TLBs \cite{tlb}). In such cases, the acquired knowledge is applied to other addresses, including previously unseen. A faithful implementation of a Markov-chain for delta transitions would be a directed graph, with deltas as states/nodes and probabilities as weighted transitions/arcs.%, connecting those nodes.

A delta is the difference between two consecutive addresses. As we can see from the simplified example below, given an initial address and a stream of deltas, the address stream can be reconstructed.
\\
\texttt{Address:\ \ \ \ \ \ 1\ \ \ 4\ \ \ 2\ \ \ 7\ \ \ 8\ \ \ 9}\\
\texttt{Delta:\ \ \ \ \ \ \ \ \ \ 3\ \ -2\ \ \ 5\ \ \ 1\ \ \ 1}
\\
In real systems, we have page limits, which constrain the reach of deltas. Both the virtual and physical memory space are divided into pages. For security and integrity reasons, the page allocation is usually not considered to be sequential. The page contents are indexed by the remaining least significant address bits and stay unaltered between translations. When prefetching, any predicted addresses that fall outside the page limits are discarded.

% THIS CAN GO TO PAGE CACHE, AND REPLACED BY A CHALLENGE FOR MARKOV CHAIN
%One challenging aspect of data prefetching is the presence of complex access patterns. A great fraction of research on prefetchers is limited to stride prefetching, which is when a single delta is used to estimate the forthcoming addresses. A rather indirect way of supporting complex patterns is the association of deltas with other information, such as the page or the instruction pointer (IP). On the other hand, distance prefetching is able to reproduce more complex patters.%, %with the expense of hardware complexity. 

%Current delta-based prefetchers, such as VLDP \cite{shevgoor15}, use a Delta Prediction Table (DPT) or equivalents, to match delta histories and predict the immediately next delta. This approach drifts away from Markov modelling, as there is no realistic notion of transition frequency. The history lookup takes place in fully-associative structures, which is not practical for a hardware implementation. This restricts the information held and impacts the prediction accuracy.

%While there are many prefetchers only supporting a single delta at a time, namely stride, 
%for recognising more complex access patterns there is a big challenge.
One challenge in distance prefetching is that many pages might be accessed in interleaving patterns and thus obfuscating the produced delta stream. The delta stream, that would otherwise be used in its entirety to update the Markov (or alternative) model, has invalidated deltas, from comparing addresses from different pages, such as
when accessing data from many sources iteratively.
%Interleaving page access patterns are very common due to the out-of-order execution in modern processors, the serialisation of accesses or the simpler case,  
%Common approaches to overcome this problem include avoiding Markov-models entirely, and/or trying to correlate the deltas to certain addresses, offsets, pages or instruction pointers (IP).
Our general idea is to track \emph{deltas per page instead of globally, but build an accurate Markov model for global decisions}. 

The main contribution of this paper is the \emph{introduction of an efficient, more-faithful representation of a Markov chain}, that provides a metric of delta-transition probability. This results in increased accuracy %and better logic and space requirements, 
for exploiting more complex access patterns.

\subsection{Motivation/ Preliminary Experiment} \label{mot}

We overview the real-world complexity of such delta-transition Markov chains, to gain an insight into related challenges and optimisations. Using a simple experiment, we monitor all the delta transitions using the competition's evaluation framework (see \ref{fr}). We implement a dummy cache prefetcher, where all occurrences of valid delta transitions (from addresses falling in the same page) are counted inside an adjacency matrix.

\begin{figure}[h!]
\includegraphics[width=0.5\textwidth , trim=5 -0 -15 0 ]{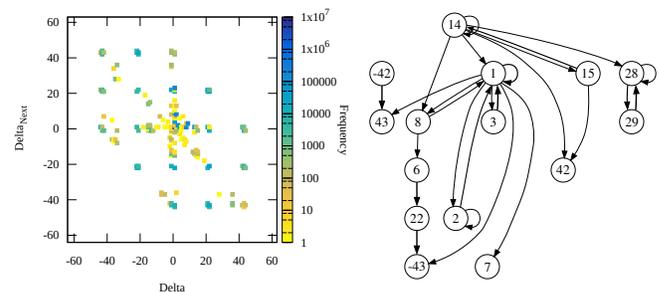}
\caption{Two visualisations for cactuBSSN}\label{mo1}
\end{figure}

Figure \ref{mo1} on the left, shows a visualisation of the frequencies for the (L2) delta transitions in a run of 607.cactuBSSN\_s-3477B. On the right, we can see the produced Markov chain (LLC), with the width of the arrows representing the probability of transition. The sum of the width of all arcs going out of a node sum to 1 (some transitions with low probability are excluded).

\begin{figure}[h!]
\includegraphics[width=0.5\textwidth , trim=11 0 -50 20 ]{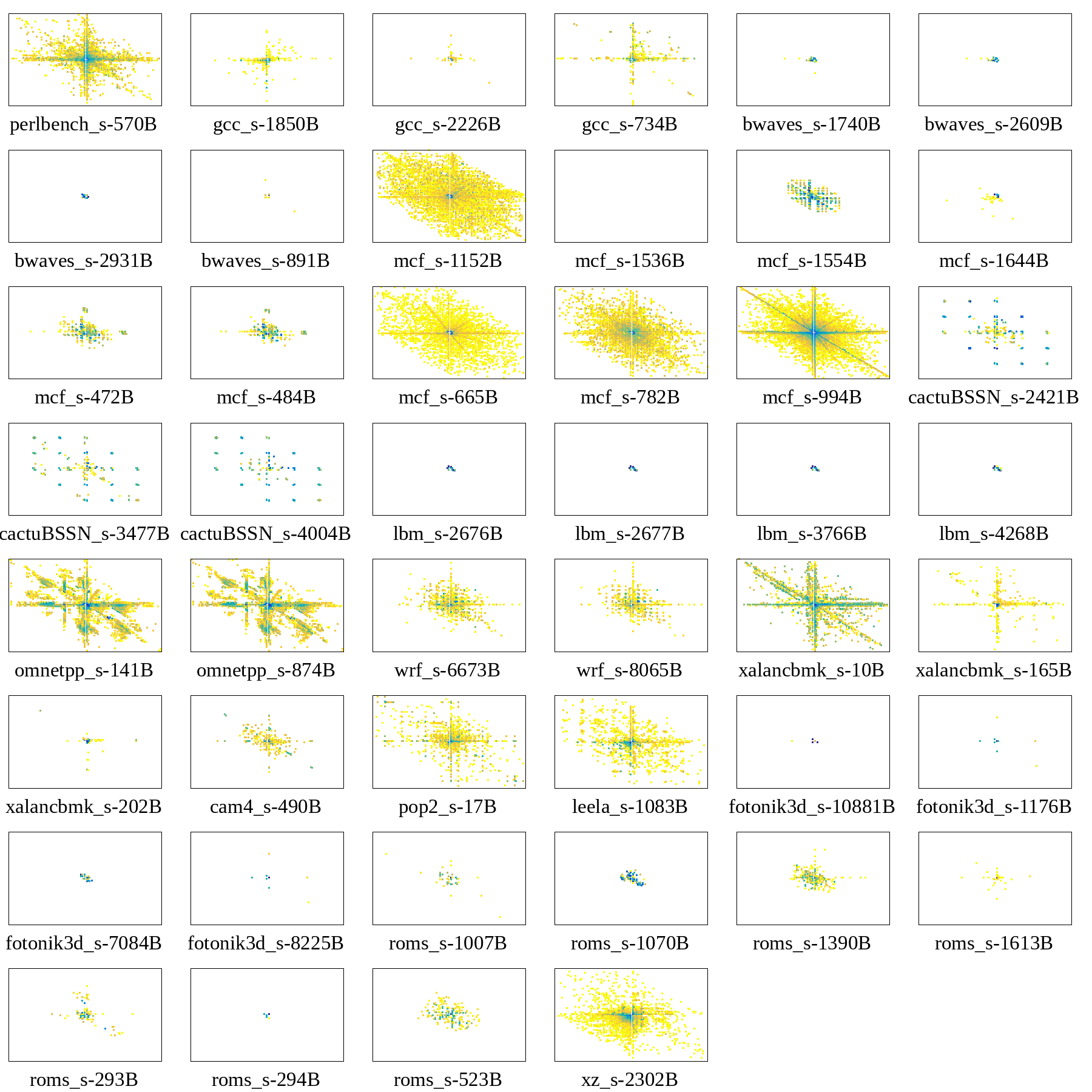}
\caption{Adjacency matrix visualisations for delta-transition frequencies}\label{mo2}
\end{figure}

Figure \ref{mo2} shows the respective visualisation of the (L2) adjacency matrx for all benchmark traces. % (same axes). 
There are some interesting observations: 1) The matrices are sparse, but 2) not as sparse to justify only supporting regular strides (such as \((1, 1)\), i.e. the model of the next line/sequential prefetcher). 3) Instead of only supporting a limited coverage of deltas \cite{michaud}, it seems worthwhile to be unbiased, including negative deltas \cite{grannaes2011storage} as well. 4) Matrices that are too sparse or empty (mcf\_s1536B), indicate simple patterns or invalidated deltas (see \ref{pca}).

Some additional observations: 1) The diagonal lines are most likely from cases of transitions from seemingly-random accesses inside a page, while a regular stride is performed. For example, in a streaming operation with a delta transition \((\delta, \delta)\), any secondary accesses would yield transitions of the form \((\delta', -\delta'+\delta)\), where \(\delta\) is the stride and \(\delta'\) is a new temporary delta. %This conclusion is a generalisation of similar patterns found in related work.
2) This also explains any vertical and horizontal lines near the axes, as such transitions are preceded and succeeded by the points  \((\delta, \delta')\) and \((-\delta'+\delta, \delta)\) respectively.
3) The reason that there seem to be `inner' bounds that make the overall shape seem like a hexagon %, instead of a square (-63 to 63 for each axis) 
is because such outliers would imply two consecutive deltas pointing outside the page margins.% (4KB or 64 lines).

%\section{Related work} 

%\section{Motivation} %(experiments, even use gephi)

\section{Proposed solution} %Prefetcher design, L2 & LLC prefetcher, L1D prefetcher}

%Divide everything in training and decisions?

\subsection {Delta cache: A novel Markov chain representation}

The main structure is an efficient representation of Markov chain for distance prefetching. %For distance prefetching, the states represent the deltas.%, instead of the addresses themselves \cite{ghb}. 
One challenge in implementing a Markov chain in hardware is that a naive accurate implementation would require N*N positions, where N is the number of states, for maintaining the transition probabilities in an adjacency matrix. For this reason and the fact that %in our case 
it usually has a high sparsity (see \ref{mot}), existing implementations approximate it with associative structures. Those associative structures (such as a fully-associative or set-associative cache) usually employ a Least Recently Used (LRU) \cite{tlb} (or approximations \cite{shevgoor15}), or a First-In First-Out (FIFO) replacement policy, which are both prone to losing track of important transitions due to thrashing. Moreover, with the information kept by LRU and FIFO, there is no real metric of frequency/probabilities, which is what Markov-chains are originally supposed to provide.

\begin{figure}[h!]
\includegraphics[width=0.28\textwidth , trim=-80 10 80 0 ]{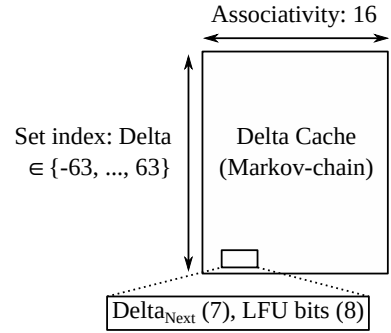}
\caption{Delta Cache (in L2 prefetcher) }\label{fig1}
\end{figure}

In figure \ref{fig1}, we present our Markov chain representation for the level 2 prefetcher. It is a set-associative cache, providing delta transitions based on the current delta. It is indexed by the current delta, and the blocks in each set represent the most frequent immediately-next deltas. Assuming that we observe line-addresses (L2 prefetcher in the framework), there are 64 possible positions (offsets) in a 4KB page. This totals in a delta size of 7 bits, representing values from -64 (excluding, since it points to a different page) to +63.  

With respect to the replacement, we use an approach similar to the Least Frequently Used (LFU) replacement policy, with the goal to keep the correct transition probabilities, but also give the opportunity of phased-out prominent deltas to be evicted quickly (slight resemblance in \cite{decay}). Each block in a set contains the next delta alongside a counter (LFU bits). This counter is incremented each time there is a hit. When there is an overflow, all blocks in the respective set have their counter values halved. In this way, we retain almost the same count proportions, with the reduced accuracy favouring higher values. In order to find the transition probability, we divide the value with the sum of all values in the set, which can also be calculated progressively.
 %kept efficiently in a counter per set (not in the evaluation).
Keeping the proportions is important in both replacement and %decision making.  for 
prioritising prefetches.

%Another challenge with approximating a realistic 

\subsection {Page cache: Reconstructing obfuscated delta transitions} \label{pca}

In this subsection we describe a mechanism designed to help reconstruct delta transitions obfuscated by `unexpected', sometimes temporary, page transitions. There are similar approaches in related work \cite{kpcp, shevgoor15}. % occurring from the serialisation of accesses, the out-of-order execution or the complexity of code. Page transitions also happen naturally from reaching the page limits.
This does not modify the decision part of Pangloss, as it only helps to increase the number of valid observations for updating the Delta cache.

%We do not modify the decision part of the prefetcher, as this only helps to increase the number of valid observations for updating the Delta cache. The values inside the delta cache are used globally, i.e. for all pages and instruction pointer (IP) values.

In a naive Markov-chain distance prefetcher, the deltas would be calculated by comparing with the latest address. Instead, we modify the algorithm to keep the last address/delta \underline{per page}, thus maximising the probability of yielding a valid transition (i.e. inside the page limits). %The last deltas per page are also kept alongside the respective last address offset, as they are required to provide delta transitions. 
%In figure \ref{fig2}, we can see an overview of the page cache.

\begin{figure}[h!]
\includegraphics[width=0.5\textwidth , trim=-15 0 -25 0 ]{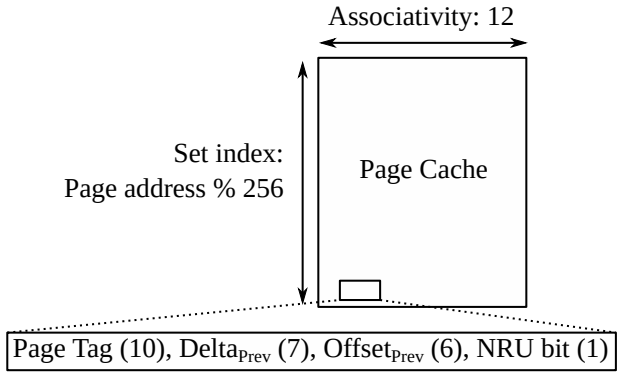}
\caption{Page Cache (in L2 prefetcher) }\label{fig2}
\end{figure}
\begin{figure*}[h]
\includegraphics[width=\textwidth , trim=30 18 10 10 ]{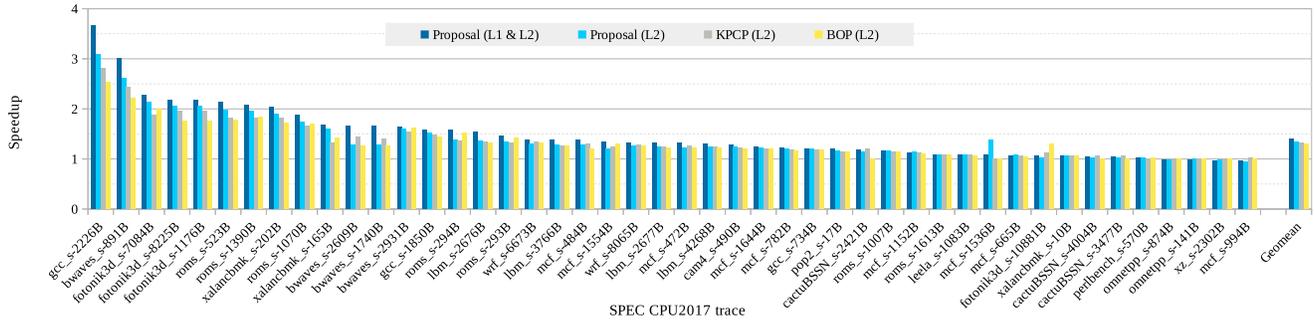}
\caption{Single-program evaluation: speedups over non-prefetch}\label{sp1}
\end{figure*}

In figure \ref{fig2}, we can see an overview of the page cache. The page cache is set-associative, indexed by the page address. Each block has 4 fields:
\begin{itemize}
\item Page tag: to identify the page and distinguish from others in the set. We found that restricting it to only 10 bits had a marginal impact on performance, despite the small probability of false positives.
\item Delta\textsubscript{Prev}: the previous delta, with which the transition is found. In the L2 prefetcher (cache line address granularity), the deltas are 7 bits long. On insertion, the value -64 is used as an initial value, to indicate that there was no previous delta, since it always points to a different page.
\item Offset\textsubscript{Prev}: the previous address offset. This is used to calculate the current delta based on the new address. %Since the rest of the address would be the page, assuming that we had a page match, only the offset is needed. 
It consumes 6 bits (values from 0 to 63).
\item NRU bit: This bit is used for approximating the LRU replacement policy with 1 bit  \cite{shevgoor15}, by always evicting the Not-Recently Used (NRU) block. %REFERENCE
\end{itemize}

\subsection {Markov-chain traversal}
Given a prefetch degree and the current delta, the prefetcher must decide how to traverse the approximated Markov chain, to provide the most profitable next deltas. Since the degree can allow paths of length \(>1\), accurately evaluating the probabilities of all possible paths in the graph becomes expensive. This is because it would require \(degree-1\) matrix multiplications involving the adjacency matrix. 

We propose a simple heuristic to predict the most likely next deltas: recursively, prefetch the addresses occurring from the child deltas with probability \(>1/3\) and proceed with the highest probable delta for the next iteration, until we count as many prefetches as the prefetch degree.

Note that if a resulting prefetch address falls out of the current page, it is discarded, % and not counted as a prefetch, 
but the path remains valid. This is done to preserve subsequent accesses to the same page, even if the same pattern started from other offsets during training.

%\paragraph {Adaptive prefetch degree (?)}

\section{Evaluation}\label{eva}
\subsection{Framework and Baseline configurations}\label{fr}% small diagram?

We are using the ChampSim micro-architectural simulator for the competition's baseline configurations for 1-core and 4-core simulations. The warmup phase takes 50M instructions and the simulation runs for another 200M. We are using the provided selection of SPEC CPU2017 benchmark traces (with over 1 Misses per K instructions (MPKI)). All runs use the same branch predictor (hashed perceptron) and cache replacement algorithm (LRU).

We compare the performance of our prefetcher to two state-of-the art prefetchers, the Best-Offset Prefetcher \cite{michaud} (BOP) and the prefetcher from KPC \cite{kpcp} (KPCP). The first was the winner of the previous Data Prefetching Championship (DPC2) and was ported to work as an L2 prefetcher in the current version of ChampSim. The latter is already included in the ChampSim repository and represents the prefetcher part of KPC.

Our final multi-level prefetcher, \emph{`Proposal L1\&L2'}, includes two versions of the same design, one for L1 and one for L2. In order to be fair with %the comparison with 
the related work, we also report  results for the single-level prefetcher, \emph{`Proposal L2'}, which is the L2 part in standalone. 

The L1 part had some additional changes to benefit from the fact that the framework allows byte-address granularity for L1. %, instead of line-address granularity for L2. 
We observed a 64-bit alignment in L1, which resulted in 512 possible offsets in a 4KB page. This increases the number of sets in delta cache to 1024, the offset size to 9 bits and the delta size to 10 bits. The LFU bits are reduced to 7.%There were some smaller details in our design space exploration, %such as the number of LFU bits and the prefetch degree, 
%which are excluded from text.

\subsection{Single-program}
Figure \ref{sp1} illustrates the single-program evaluation. Our solution achieves a geometric speedup of 6.8\% and 8.4\% over KPCP and BOP respectively. The geometric speedup over non-prefetch is 40.4\%. The respective geometric speedups for the single-level version are 1.7\%, 3.2\% and 33.5\%.

\subsection{Multi-program}

In order to produce representative program mixes, we divide the 46 traces in two groups, the `low' and `high', for those yielded a speedup of 1.3  and below (last 21 from fig. \ref{sp1}) and those above 1.3 respectively, using `Proposal L1 \& L2'. Then, for each of the 5 group combinations (l-l-l-l, l-l-l-h, ...) we produce 8 random mixes, totalling 40 mixes.

\begin{figure}[h!]
\includegraphics[width=0.5\textwidth , trim=-5 10 -35 10 ]{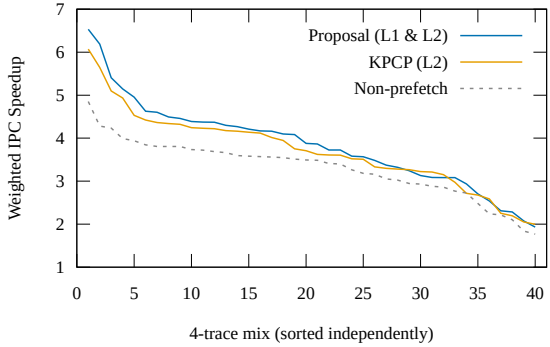}
\caption{Multi-program evaluation%: weighted IPC speedups of 40 representative 4-program mixes, over when running in isolation% (single-core, non-prefetch)
}\label{sp2}
\end{figure}

Figure \ref{sp2} shows the weighted IPC speedups  of `Proposal L1 \& L2' and KPCP over the single-core runs with non-prefetch (i.e. \( \sum (IPC_{i}/IPC_{alone\_i})\) \cite{cruise}). It is clear that the multi-level prefetcher performs generally better than KPCP, while the single-level version (not shown) is roughly in-between. %If we sum up all individual IPCs (i.e. throughput \cite{cruise}), the speedup of our proposal is 1.8\% over KPCP. 

%\vspace{-3em}

\subsection{Resources}
\paragraph{Space budget}

In table \ref{tab1}, we present the space requirements of our multi-level prefetcher. The total number of bits (59.4 KB) is below the space budget of the competition (64 KB) for the single-core configuration. Since we have not included an LLC prefetcher, the space requirements for the multi-core configuration is multiplied by 4, which is also under the competition's space budget (4\(\times\)64 KB). In the single-level case, the L2 prefetcher only consumes 13.1 KB.

\begin{table}[h!] 
\footnotesize
%\scriptsize
\centering
   
\hskip-0.25cm
\begin{tabular} {l| c c }
%\hline
&Description (bits)&(KB)\\
\hline
L1D:&&\\%\((46.3)\)\\
Delta cache& \(1024\ sets \times 16\ ways \times (10 + 7)\) &\(34.8\)\\
Page cache& \(256\ sets \times 12\ ways \times (10 + 10 + 9 + 1)\) &\(11.5\)\\
\hline
L2:&&\\%\((13.1)\)\\
Delta cache& \(128\ sets \times 16\ ways \times (7 + 8)\) & \(3.8\) \\
Page cache&\(256\ sets \times 12\ ways \times (10 + 7 + 6 + 1)\)&\(9.2\)\\
\hline

LLC:&\emph{None}&\(0.0\)\\
\hline
Total&&\(59.4\)\\
%\hline

%\hline
\end{tabular}

\begin{tabular}{c}
\\
\end{tabular}
\caption{Single-core configuration budget} 
\label{tab1}
\vspace{-1em}
\end{table}

\paragraph{Logic complexity}

Pangloss is H/W-friendly. %was designed to be lightweight. %In contrast to popular related work, o
%Our structures are set-associative instead of fully-associative caches. 
The low associativity in the Page Cache and Delta Cache %(up to 16)
ensures that there will be few simultaneous comparisons of few bits. This allows keeping more information in a concise space. The Markov traversal heuristic that selects probabilities above 1/3, implies that up to 2 child deltas will be selected. Thus, %for prioritising and proceeding with longer delta paths, 
one extra comparison is enough to point to the next layer. %In order to find the transition probabilities, we also need the sum of all counters in the set, which can be kept efficiently in a counter per set.
%. This can be done efficiently by keeping an individual sum counter for each set. 
For a medium prefetch degree, the recursive lookup \cite{shevgoor15} remains relatively efficient, although allowing a delay could also prove beneficial for timeliness \cite{michaud}.
%In order to find the top 1 out of the 16 for prioritising and proceeding with longer delta paths, we only need a single comparison of the filtered 2, which  happen to be the top 2. 

According to the use case, many parameters that impact the space/logic complexity, can be explored further.

\section{Future work and Conclusion}

%Since prefetching in hardware is a prediction mechanism, it is difficult to have an optimal data prefetcher, but it is rewarding to improve. 
%Each prefetcher is a prediction mechanism, having its own weaknesses. 
All prediction mechanisms have some weaknesses. 
When observing short repeating delta patterns, such as 1, 1, 2, 1, 3, 1, 1, 2, 1, 3, ..., the transitions (1, 1), (1, 2) and (1, 3) would yield an equal probability. This in combination with other factors, like a low prefetch degree, could have a performance overhead. This does not happen with multiple-delta histories \cite{shevgoor15}. However, multiple-delta history matching could be negatively affected by some memory hierarchy effects that reorder or even hide deltas. Systematically evaluating the probability and overhead of pattern conflicts in Pangloss would be desirable. Alternatively, we could evaluate the presence of multiple-delta states in the Markov chain. One mechanism that differs from random walks on a Markov-chain is the traversal heuristic, which can be explored further. %Finally, we could also experiment with adding features from other prefetchers, such as the delay queue from the best-offset prefetcher \cite{michaud}.

%In this paper, we introduce a prefetching scheme able hold more information efficiently and predict more complex patterns than the selected state-of-the-art baselines. 

In this paper, we introduce a H/W-friendly prefetcher with a more-faithful representation of a Markov chain, resulting in a higher accuracy and performance.
 %able hold more information efficiently and predict more complex patterns than the selected state-of-the-art baselines. 

%REMOVE?
%It could prove useful to combine with a delay queue \cite{michaud}. % with expected times being kept inside our Page/Delta caches. 
%Also, the Markov-chain traversal heuristic here could be explored further. Another area for improvement comes from the fact that reaching outside a page is forbidden. However, there could be a buffer to remember page transitions to help predict successive pages. 
%Finally, although we opted-out from using IPs for delta correlation, it might worth investigating a branch-predictor/data-prefetcher combination, for accurately predicting next IPs and any correlated deltas.

%We present a ... In out test framework, Pangloss achieves a geometric speedup of 6.8\% over a state-of-the-art
 
%and attempt to overcome any context-switching-related complication
% overhead of such pattern conflicts inside the Markov model. %This would balance prefetching for different workloads out, and become immune to.

\section*{Acknowledgement}
This research was sponsored by dunnhumby. The support of the United Kingdom EPSRC (grant numbers EP/L016796/1, EP/N031768/1, EP/P010040/1 and EP/L00058X/1) is gratefully acknowledged.
%%%%%%% -- PAPER CONTENT ENDS -- %%%%%%%%

%%%%%%%%% -- BIB STYLE AND FILE -- %%%%%%%%
%\bibliographystyle{ieeetr}
%\bibliography{ref}

%\AtNextBibliography{\fontsize{9.1}{10}\selectfont}
%\AtNextBibliography{\fontsize{9}{9}\selectfont}
%\AtNextBibliography{\fontsize{8.3}{7.95}\selectfont}
%\AtNextBibliography{\fontsize{8}{7.9}\selectfont}
\AtNextBibliography{\fontsize{9}{9}\selectfont}

\printbibliography

\vspace{12pt}

%%%%%%%%%%%%%%%%%%%%%%%%%%%%%%%%%%%%

\end{document}